\documentclass[twocolumn,prl,showpacs,preprintnumbers,superscriptaddress,amsmath,amssymb,citeautoscript]{revtex4-1}
\usepackage{graphicx}
\usepackage{epstopdf}
\usepackage{dcolumn}
\usepackage{color}
\usepackage{bm}
\usepackage{rotating}
\graphicspath{{./figs/}}
\usepackage{multirow}
\usepackage{rotating}
\usepackage{array}

\newcommand{\bs}{\boldsymbol}
\newcommand{\tX}{\tilde{X}}

\DeclareMathOperator{\diag}{diag}
\DeclareMathOperator{\im}{Im}
\DeclareMathOperator{\re}{Re}
\DeclareMathOperator{\Li}{Li}

\begin{document}

\title{Controlling topological superconductivity by magnetization dynamics }
\author{Vardan Kaladzhyan}
\email{vardan.kaladzhyan@cea.fr}
\affiliation{Institut de Physique Th\'eorique, CEA/Saclay,
Orme des Merisiers, 91190 Gif-sur-Yvette Cedex, France}
\affiliation{Laboratoire de Physique des Solides, CNRS, Univ. Paris-Sud, Universit\'e Paris-Saclay, 91405 Orsay Cedex, France}
\author{Pascal Simon}
\email{pascal.simon@u-psud.fr}
\affiliation{Laboratoire de Physique des Solides, CNRS, Univ. Paris-Sud, Universit\'e Paris-Saclay, 91405 Orsay Cedex, France}
\author{Mircea Trif}
\email{trif@mail.tsinghua.edu.cn}
\affiliation{Institute for Interdisciplinary Information Sciences, Tsinghua University, Beijing}

\date{\today}

\begin{abstract}
We study theoretically a chain of precessing classical magnetic impurities in an $s$-wave superconductor. Utilizing a rotating wave description, we derive an effective Hamiltonian that describes the emergent Shiba band. We find that this Hamiltonian shows non-trivial topological properties, and we obtain the corresponding topological phase diagrams both numerically and analytically. We show that changing precession frequency offers a control over topological phase transitions and the emergence of Majorana bound states. We propose driving the magnetic impurities or magnetic texture into precession by means of spin-transfer torque in a spin-Hall setup, and manipulate it using spin superfluidity in the case of planar magnetic order.  
\end{abstract}


\maketitle

{\it Introduction.} The search for topological phases of matter during the last decade has led to remarkable advancements in engineering systems with preassigned exotic excitations such as the Dirac,  Weyl, or Majorana fermions. The latter have been pursued in numerous condensed matter setups \cite{Sarma2015}, as they have been suggested as promising candidates for fault-tolerant topological quantum computing \cite{Kitaev2003}. 

Ubiquitous and destructive by its nature for other phenomena, disorder has become one of the most interesting and reliable tools to build the sought-for topological systems. Discovered more than half a century ago \cite{Yu1965,Shiba1968,Rusinov1969,Sakurai1970} impurity-induced bound states in superconductors have been recently brought to life in the experiments \cite{Yazdani1997,Menard2015}. The latter, along with the rise of topological phases of matter, initiated a series of works, both theoretical \cite{Choy2011,Nadj-Perge2013,Nakosai2013,Braunecker2013,*Klinovaja2013,*Vazifeh2013,Pientka2013,Pientka2014,Poyhonen2014,Heimes2014,Ojanen2015,Rontynen2015,Brydon2015,Peng2015,Braunecker2015,Zhang2016,Hoffman2016,Neupert2016,Kimme2016,Kaladzhyan2016a} and experimental \cite{Nadj-Perge2014,Ruby2015a,Pawlak2016,Feldman2017}, proposing to use Shiba states as promising building blocks for desired Majorana-supporting systems. The underlying mechanism is reminiscent of that of electronic bands appearing in solids: being brought together discrete Shiba levels originating from different impurities  hybridize and form Shiba bands, with electrons filling them according to the Pauli principle. The resulting band structure corresponds to that of a $p$-wave, or topological  superconductor, that can exhibit  Majorana edge modes depending on the parameters of the system under consideration. The drawback of such an implementation, however, is that system parameters are typically fixed, and one cannot explore easily the full phase diagram.

\begin{figure}[t]
	\includegraphics[width = 0.99\columnwidth]{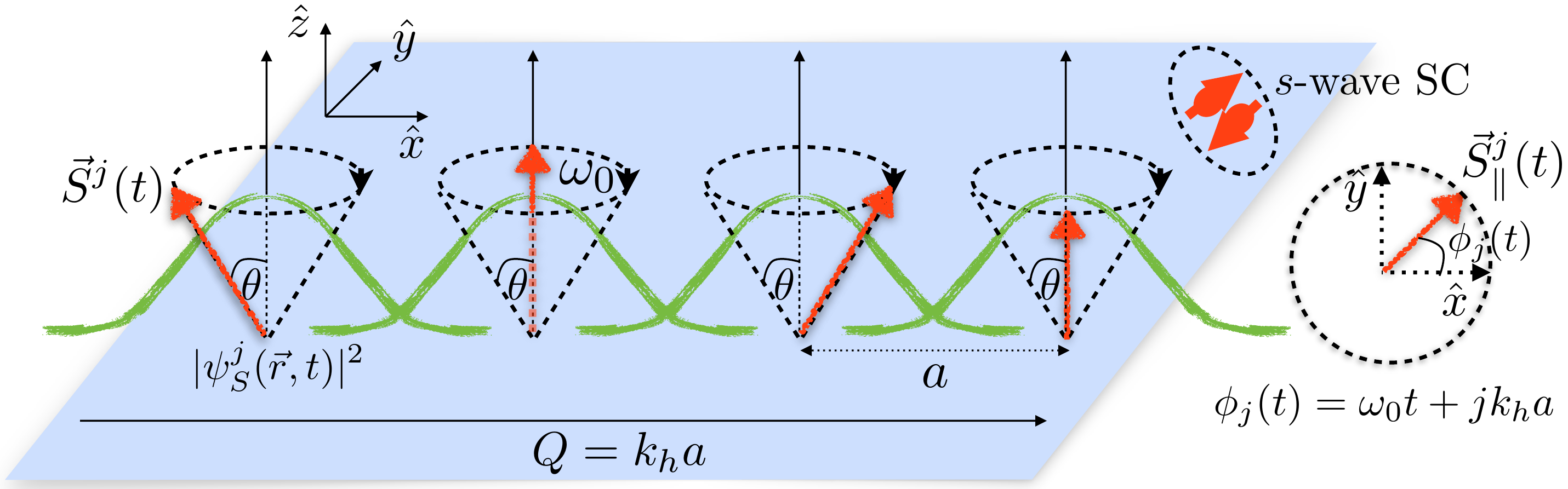}
	\caption{Sketch of the precessing spin helix in a two dimensional $s$-wave superconductor (the blue plane). The classical spins (red arrows) are separated by a distance $a$ and precess around the $z$ axis with a frequency $\omega_0$ at a polar angle $\theta$. The precession azimuthal angle $\phi_j(t)=\omega_0 t+k_hja$, with $k_h$ the step of the helix and $j$ the position of the spin in the chain. In green it is shown the local dynamical Shiba states wave-functions which overlap to give eventually a dynamical Shiba band.  }
	\label{system}
\end{figure}

In this Letter, motivated by the recent progress in the so called dynamical, or Floquet topological insulators \cite{RudnerPRB10,LidnerNatPhys11}, we present a new promising setup not only for engineering a topological superconducting phase, but most remarkably for {\it controlling} topological phase transition by means of magnetization texture dynamics. We consider theoretically a ``dynamical Shiba chain",  that pertains to a set of classical magnetic impurities with precessing spins deposited on top of a 2D s-wave superconductor (see Fig.~\ref{system}).  We find that such a dynamical magnetic texture can give rise to a non-trivial Shiba band which can be controlled by tuning the precession frequency. Such features are different from previous time-dependent Floquet superconducting systems see, for example, Refs.~[\onlinecite{JiangPRL11,*TrifPRL12,*TongPRB13,*ManishaPRB13,*PlateroPRL13,*LiuPRL13,*ViolaELP15,*LossPRL2016}], in that the band is not manipulated directly by external fields, but indirectly, by the dynamics of the magnetic texture that stirs the underneath superconductor and cause the appearance of such a band. This is inherently a strong coupling regime, as the magnetic texture is the reason for such band to occur in the first place.  

{\it Model.} The Hamiltonian describing our dynamical systems reads \cite{Nadj-Perge2013}
\begin{equation}
H_{\rm tot}(t)=H_0+H_{\rm imp}(t)\,,
\end{equation}
where
\begin{align}
H_0& = \xi_k \tau_z + \Delta_s\tau_x\,,\\
H_{imp}(t) &= \sum\limits_j {\bs J}_j(t) \cdot \bs \sigma\ \; \delta(\bs r - \bs r_j)
\end{align}
being the sum of the Bogoliubov-DeGennes Hamiltonian for the superconductor and its coupling to the magnetic impurities, respectively. Here, $H_0$ is written in the Nambu basis $\left\{c_{\bs k \uparrow}, c_{\bs k \downarrow}, c^\dag_{-\bs k \downarrow}, -c^\dag_{-\bs k \uparrow} \right\}^{\mathrm{T}}$, with $\bs \sigma = (\sigma_x, \sigma_y, \sigma_z)$ and $\bs \tau = (\tau_x, \tau_y, \tau_z)$ matrices acting in spin and particle-hole subspaces respectively. The superconducting order parameter is denoted by $\Delta_s$, the spectrum of free electrons is defined as $\xi_k \equiv k^2/2m-\varepsilon_F$, where $\varepsilon_F$ is the Fermi energy. Below we also set $\hbar$ to unity. For the periodically driven magnetic chain  we assume that the impurities are localized at positions $\bs r_j$, and have precessing spins that are defined as ${\bs J}_j(t) \equiv J \left[\sin\theta \cos (\omega_0 t +\phi_j),\, \sin\theta \sin (\omega_0 t +\phi_j),\, \cos\theta \right]$  with precession frequency $\omega_0$, polar angle $\theta$ as shown in Fig.~\ref{system}, and equidistant  individual phase shifts $\phi_j \equiv k_h a j,\, j \in \mathbb{Z}$. In the latter $a$ denotes the spacing between impurities, and $k_h$ is the so-called helix step. 

The time-dependent Schr\"odinger equation reads
$
i \partial_t \Psi\left(\bs r, t \right) = H_{\rm tot}\Psi\left(\bs r, t \right).
$
This Hamiltonian is periodic, $H_{\rm tot}(t+T)=H_{\rm tot}(t)$, with $T=2\pi/\omega_0$ and, moreover,  the symmetry of the problem allows us to perform a time-dependent unitary transformation that makes the problem fully static. We can write $\Psi\left(\bs r, t \right)=U(t)\Phi\left(\bs r\right)e^{-iEt}$, with $U(t)=e^{-i\omega_0 t \sigma_z/2}$  so that we obtain the stationary Schr\"odinger equation:
\begin{equation}
\left[H_{\rm tot}(0) - B\sigma_z\right]\Phi\left(\bs r\right) = E \Phi\left(\bs r\right)\,,
\label{SchrStat}
\end{equation}
where the fictitious magnetic field $B \equiv\omega_0/2$  is perpendicular to the plane of the superconductor, which will be referred to as `driving frequency' hereinafter, and  $E$ is the quasi-energy defined modulo $\omega_0/2$. Let us make now a more concise connection with the usual stroboscopic, or Floquet  description of periodically driven systems. The full evolution operator for the driven chain can be written as:
\begin{equation}
U_{\rm tot}(t)=e^{-iB\sigma_zt}e^{-i\mathcal{H}_{F}t}\,,
\label{EvOp}
\end{equation} 
with $\mathcal{H}_{F}\equiv H_{\rm tot}(0)-B\sigma_z$. After one period $T$, the evolution operator can be written $U_{\rm tot}(T)=\exp{(-i\mathcal{H}_{F}T)}$ (up to a sign), with $\mathcal{H}_F$ identifying as the Floquet Hamiltonian describing the evolution of the system at $t=nT$, with $n\in\mathbb{N}$ (stroboscopically).  The Hamiltonian $\mathcal{H}_{F}$ gives rise to a quasi-energy spectrum defined up to integer multiples of  $2B$ and, as in the static case, can result in non-trivial topological properties which in an open system are identified  with the appearance of edge states. However, it does not fully characterize the topological structure and the entire spectrum of edge states of the driven system. Such a complete description was  developed recently in several works, where they showed that in order to fully describe that, one needs the evolution operator at all times $t$, not only at $t=T$. However, for our situation of circular spin texture precession, it turns out that $\mathcal{H}_{F}$ describes fully the topological structure of the driven system, and thus we focus on that aspect only in the following. 

As discussed in Ref.~\cite{Kaladzhyan2016c} a single magnetic impurity with a periodically driven spin gives rise to a pair of Shiba states residing in the effective gap $\Delta_s^{\rm eff} = \Delta_s - B$, provided the driving frequency $B$ is smaller than the superconducting gap $\Delta_s$. This condition is essential to have a gapful system and well-defined impurity-induced subgap states. The energies of these states in the deep-dilute regime ($\alpha \sim 1$) are given by $\pm \epsilon_0(B)$, where 
\begin{equation}
\epsilon_0(B) \equiv \left[\left(1 - \frac{1}{\alpha}\right) \Delta_s - B \cos \theta\right]\,,
\end{equation}
and $\alpha \equiv \pi \nu_0 J$ is the dimensionless impurity strength parameter written in terms of normal-phase density of states $\nu_0$. It has been shown in Refs.~\cite{Pientka2013,Pientka2014} that a static helical chain of magnetic impurities produces a $4\times4$ Shiba band structure with non-trivial topological properties. Moreover, for $\alpha\approx1$ one can project the resulting $4\times4$ Hamiltonian onto an effective  $2\times2$ that fully characterize the low-energy spectrum (the energy separation between the bands is of order $\Delta_s$).  

Hereafter we use Eq.~(\ref{SchrStat}) and, following the procedure described in Ref.~\cite{Pientka2013}, we derive the effective $2\times2$ Hamiltonian for the emerging Shiba band. The details of this derivation are given in the Supplementary Material (SM) \cite{sm}. 

{\it Effective band structure.}|The effective Hamiltonian describing the Shiba band in the rotating frame in both aforementioned cases can be written exploiting the $\bs d$-vector notation as
\begin{equation}
\mathcal{H}_{\rm S}(k) = d_0(k) + \bs{d}(k)\cdot \bs{\Sigma}\,,
\label{Hamildvect}
\end{equation}
with 
\begin{align}
d_0(k) &=   \left[\Delta_s \cos\theta - B(1-\alpha \sin^2\theta) \right]F_0(B, ka, k_Fa)\nonumber\,,\\
d_x(k) &=   \left(\Delta_s - \alpha B \cos\theta\right)F_x(B, ka, k_Fa)\sin\theta\nonumber\,, \\
d_z(k)  &=  -\epsilon_0(B) + \left(\Delta_s -  B \cos\theta\right)F_z(B, ka, k_Fa)\,,
\label{vectors}
\end{align}
and $d_y(k)\equiv0$.  Eqs~\eqref{vectors}  represent one of our main results.
Here $\bm{\Sigma}=(\Sigma_x,\Sigma_y,\Sigma_z)$ represents a resulting Nambu space which, however, is a complicated admixture of $\bm{\sigma}$ and $\bm{\tau}$.  The form of the functions $F_{0,x,z}(B, ka, k_Fa)$ is in general too complicated to be displayed. However, there are various limiting cases where analytical progress is possible. In this paper we focus on two limiting cases that can be studied both analytically and numerically,  i.e. the short and the long coherence length, respectively. The first case corresponds to a chain with only nearest neighbor hopping, in other words, the case of a small coherence length $\xi \ll a$, where $\xi \equiv v_F/\sqrt{\Delta_s^2-B^2}$. In this limit, we need to set in Eq.~(\ref{vectors}) the following functions: 
\begin{eqnarray}
F_{0,x}(B, ka, k_Fa)&\equiv &  \tilde{X}_{0,1}(a) \sin \frac{k_h a}{2} \sin ka\,,\\\nonumber
F_z(B, ka, k_Fa)  &\equiv &\tilde{X}_0(a) \cos \frac{k_h a}{2} \cos ka,
\end{eqnarray}
where
\begin{eqnarray*}
\tilde{X}_{0(1)}(a) &=& -\frac{2}{\pi}\, \mathrm{Im}(\mathrm{Re})\, \mathrm{K}_0 \negthickspace\left[-i\left(1+i\frac{1}{k_F \xi} \right) k_F a\right]\,,
\end{eqnarray*}
with $k_F$ being the Fermi momentum and $\mathrm{K}_0$ denoting the zeroth modified Bessel function of the second kind (for further details see \cite{sm} as well as \cite{Kaladzhyan2016b}). Note that the functions $\tilde{X}_{0,1}$ depend at least quadratically on the fictitious magnetic field  $B$, and for $B\ll\Delta_s$ we can neglect such dependence in leading order . 

The second limiting case describes a chain with very extended Shiba states, i.e. with large coherence length compared to the impurity spacing, $\xi \gg a$. Contrary to the small coherence length regime, here all the higher order hopping processes become possible. In this regime we obtain the following expressions for the functions $F_{0,x,z}$ in  Eq.~(\ref{Hamildvect}) $F_{0,x} \equiv\left[ F_{0,1}^-(k) - F_{0,1}^+(k)\right]/2$ and $F_z\equiv \left[ F_0^-(k) + F_0^+(k)\right]/2$, where we defined
\begin{align*}
F^s_{0(1)}(k) \equiv \sqrt{\frac{2}{\pi k_F a}} \mathrm{Im}(\mathrm{Re})\, f_s(k),\;
\end{align*}
with $s = \pm$ and
\begin{align}
f_s(k)&= e^{-i\frac{\pi}{4}} \left[\mathrm{Li}_{\frac{1}{2}}\left( e^{ i(k+s k_h/2-k_F)a} \right)\right.\nonumber\\
&\left.+\mathrm{Li}_{\frac{1}{2}}\left( e^{- i(k+ s k_h/2+k_F)a} \right) \right]\,,
\end{align}
expressed in terms of the polylogarithm function $\mathrm{Li}(x)$.  

Note that $d_x(k)$ in the expressions given above plays the role of the gap parameter $\Delta_k$ from Ref.~\cite{Pientka2013}, which, in the limit of $B\ll\Delta_s$ is only slightly reduced by the fictitious field. On the other hand, $d_z(k)$ is strongly affected by the driving, as it results in a shift of the alignment of the Shiba bands, and eventually their topology. While $d_0(k)$ does not change the topology of the bands, it does affect their overlap (the absolute gap), and it can also depend strongly on $B$ for $\theta\rightarrow\pi/2$ (planar helix). In fact, in such a case, the entire dependence on the magnetic field arises through this term in leading order which, however, is small for $\alpha\sim1$. 

\begin{figure}
	\includegraphics*[width = 0.99\columnwidth]{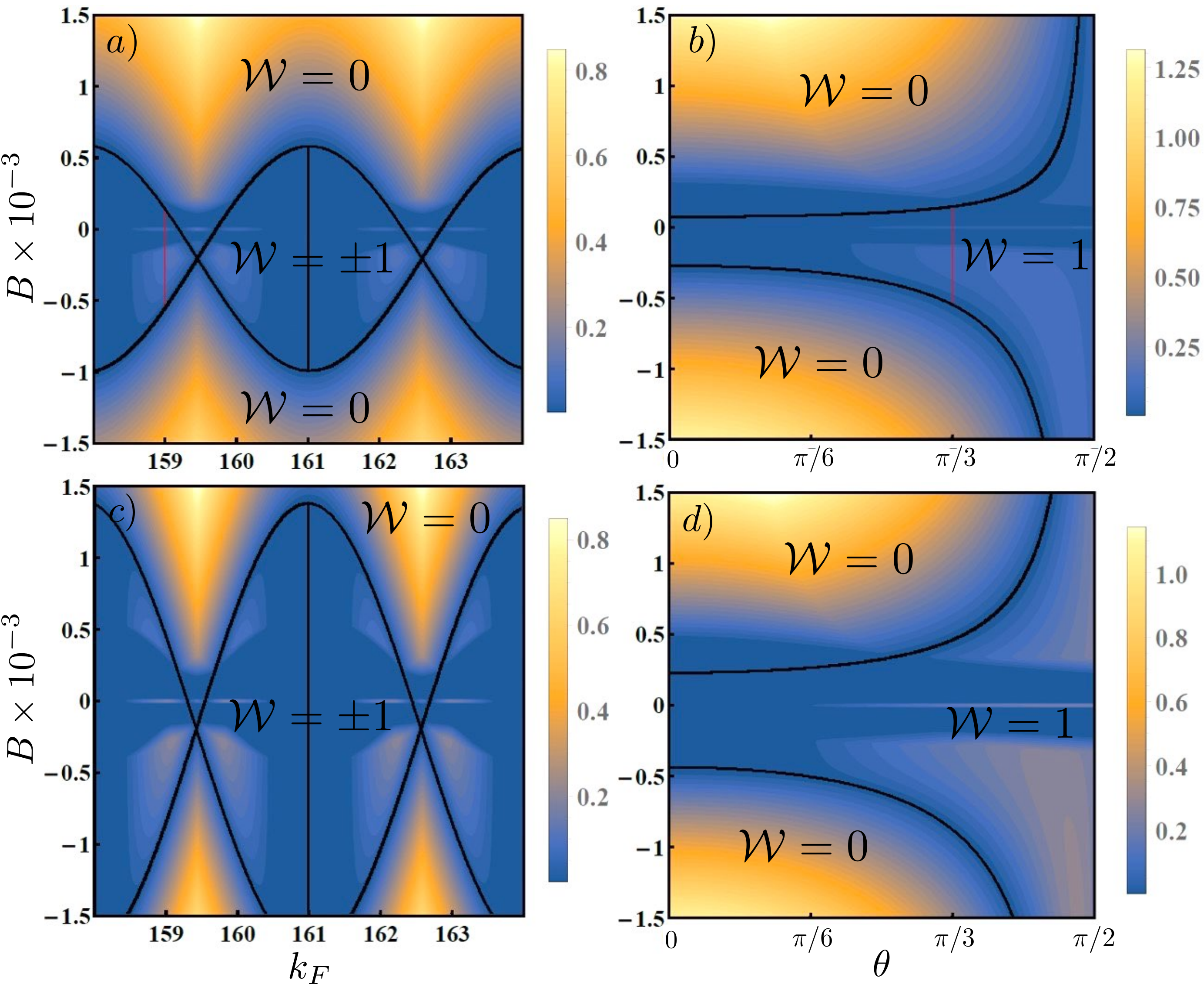}
	\caption{(color online) The gap around quasienergy $E=0$ and the winding number of the Shiba band for the small and large coherence length regimes (first and second rows respectively), plotted as functions of the driving frequency $B$ and the Fermi momentum $k_F$ (precession angle $\theta$) in the left (right) column. The continuos black lines separate regions with different winding numbers $\mathcal{W}$ which are well defined even for the gapless regions. The vertical red lines highlight the existence of localized Majorana end states in an open system [see Fig.~\ref{TBSpectrum} for details]. We set $k_h=\pi/4, v_F = 0.2, \Delta_s = 1, a = 1, \alpha = 0.9999$. The polar angle $\theta=\pi/3$ and $k_F = 159$ in the left and right columns correspondingly.}
	\label{GapWithB}
\end{figure}

{\it Quasi-spectrum and topology.} In what follows we study the topological properties of the Hamiltonian in Eq.~(\ref{Hamildvect}) in the short and long coherence length regimes introduced above. The spectrum can be found easily as $E(k)=d_0(k)\pm\sqrt{d_z^2(k)+d_x^2(k)}$ which, because of the periodic drive, is uniquely defined only up to an integer multiple of $B$. Thus, we need to fold the resulting spectrum into the first quasi-energy Brillouin zone, $E(k)\in[-B, B)$. The resulting one-dimensional Hamiltonian is real, and thus it belongs to the BDI symmetry class \cite{RyuPRB10}. In this case the number of Majorana states emerging at one end in the case of open boundary conditions is given not by a $\mathbb{Z}_2$, but by a $\mathbb{Z}$ invariant \cite{TewariPRL12}, which reads: 
\begin{equation}
\mathcal{W}=\frac{1}{2\pi}\int_{-\pi}^\pi d\theta(k)\,,
\label{winding}
\end{equation}
with $\theta(k)={\rm Arg}[d_x(k)+id_z(k)]$. This winding number characterizes the number of edge states. However, it does not indicate the presence of an absolute gap in the system, meaning that it can be well defined even if the system is gapless. We  depict such surprising features in Fig.~\ref{GapWithB}, where we plot the absolute gap between the Shiba bands, as well as the corresponding winding number, as functions of the driving frequency $B$ against the Fermi momentum $k_F$ (angle $\theta$) in the left (right) column. Both the driving frequency and the precession angles are tunable parameters and, most strikingly, this shows that the system can undergo a topological phase transition by changing the driving frequency. 

We note that for the small coherence length regime (top row) the winding number can be calculated analytically (see \cite{sm}), whereas for the large coherence length (bottom row) we restrict ourselves to computing the integral in Eq.~(\ref{winding}) only numerically \footnote{It is worth mentioning that the topologically nontrivial regions in Fig.~\ref{GapWithB} can be also determined by utilizing Pfaffian invariants reflecting parity of the winding number $\mathcal{W}$, as well as by employing the so-called `singular points technique' developed in Ref.~\cite{Kaladzhyan2016SP}}. A few more comments are in order.  As expected for $\theta = 0$ (corresponding to a ferromagnetic arrangement of the impurity spins) the gap is absent and  the system is in a gapless trivial phase with zero winding number. Conversely, when $\theta = \pi/2$ the spin helix is planar. This in turn means that the fictitious magnetic field $B$ appearing in the rotating frame [see Eq.~(\ref{SchrStat})] does not couple to the chain, which explains why  no change of phase occurs while changing the driving frequency for $\theta=\pi/2$. Therefore, the system always enters a topological superconducting phase.

\begin{figure}
	\includegraphics*[width = 0.99\columnwidth]{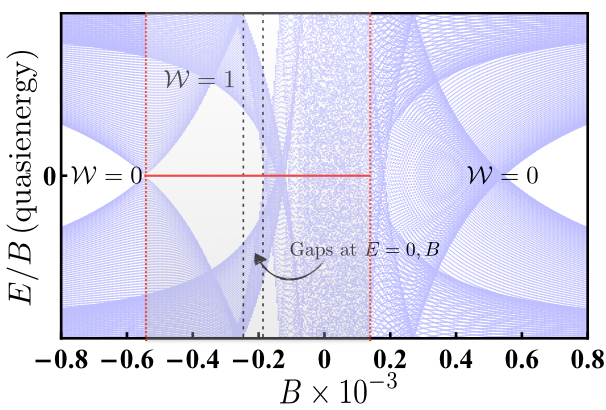}
	\caption{(color online) The quasi-spectrum (normalized by the driving $B$) for an open Shiba chain, in the regime of  small coherence length as function of the driving frequency $B$.  The horizontal red line stands for the zero quasi-energy Majorana end mode, while $\mathcal{W}$ defines the bulk winding number (see main text). A region with two gaps, at $E=0$ and $E=B$ exist, but the latter is trivial as we find no Majoranas emerging. We set $k_F=159, k_h=\pi/4, v_F = 0.2, \Delta_s = 1, a = 1, \theta=\pi/3, \alpha = 0.9999$.}
	\label{TBSpectrum}
\end{figure}

One of the most important signatures of topological systems are topological edge states. In Fig.~\ref{TBSpectrum} we show the quasi-spectrum for a dynamical chain with open boundary conditions and for the case of a short coherence length. We see that  Majorana bound states (MBS) emerges at zero energy (red line), and that their existence range is in perfect agreement with the bulk winding number calculation. Moreover, we found that the MBS even exist  in regions where the system is  gapless, albeit they are not protected anymore by the gap and any impurities could easily mix them with the bulk (extended) states. While for a region of the parameter space we found gaps at both $E=0$ and $E=B$ (see Fig.~\ref{TBSpectrum}), only the modes at the former are emerging for the circular driving utilized in this paper. However, such a conclusion should not hold for more general drivings of the magnetic texture.  

{\it Detection and physical implementations.} The dynamically generated  MBS described above could be detected in transport measurements by nearby voltage biased STM tip \cite{KunduPRL13}. Alternatively, one could utilize a recent scheme that relies on the pumped charge by the precessing texture into the STM tip at different position in the chain in the absence of any applied voltage \cite{Kaladzhyan2016c,trifUnpublished}. In order to generate the dynamics,  we envision several implementations, depending on the way the magnetic texture emerges in the first place. In the case of a pre-formed helix, either due to the the RKKY interaction mediated by electrons in the superconductor \cite{Braunecker2013,Klinovaja2013,Vazifeh2013}, or due to the SOI in the substrate \cite{LiNatCom16}, the precession of the helix corresponds simply to  global rotations. The traditional way to excite such a mode is by driving the helix with microwaves that excite the ferromagnetic resonance associated with such a rotation mode.  However, in recent years there have been tremendous progress in exciting magnetic devices in transport setup by means of the spin Hall effect \cite{SinovaRMP15}. Such a setup would allow for an all-electrical implementation of a dynamical magnetic texture - superconductor hybrid, with a controllable frequency (see \cite{sm} for details on the implementation). Both these methods can give rise to rotations of the helix, but do not result in changes of the pitch.  However, when  the impurities form a planar ferromagnet (with exchange interaction keeping the spins in a plane), it becomes possible to  control the pitch $k_h$, the frequency $\omega_0$, and the cone angle $\theta$ by means of spin biases, as showed recently in several works \cite{KonigPRL01,*SoninAdvMat10,*TakeiPRL14}. This goes by the name of spin superfluidity, as there is a direct mapping between a superfluid flow (such as in He$_4$) and the magnetization flow in such a planar spin configuration. 
As detailed in
\cite{sm}, such manipulations are possible simply by changing the spin biases induced by the spin Hall effect applied over the planar spin configuration, with a pitch in one-to-one correspondence with the spin super-current flowing through the magnetic system, and an adjustable frequency depending on the relative biases \cite{SoninAdvMat10,TakeiPRL14}. 

{\it Discussions and perspectives.} The setup proposed in this work can be generalized to  a chain of precessing magnetic impurities deposited on top of a 3D superconductor. Despite a modification in the Shiba wavefunction coherence length, we expect no qualitative difference in our main argument concerning a controlled topological phase transition. Moreover, a 3D superconductor is expected to reflect the short coherence length regime, whereas a 2D one -- the long coherence length regime. As a future extension of this work we propose to consider more complicated networks of driven magnetic impurities, e.g. a 2D array. Also, generalizations to more complicated textures and precessions is in order, as our perfect rotation wave description would break down, and a fully Floquet approach would be required. The same arguments should apply when the substrate (superconductor) posses spin-orbit interaction.

In conclusion, in this paper we proposed a way to engineer a controllable topological phase transition by means of magnetization texture dynamics. We have shown that a chain of precessing classical spins deposited on top of an s-wave superconductor gives rise to a topologically non-trivial Shiba band, and we have demonstrated that topological phase transitions in such a band can be controlled by changing the driving frequency,  a tunable parameter in the spin transport experiments.  

{\it Acknowledgements.} We would like to thank  Cristina Bena, So Takei and Yaroslav Tserkovnyak  for useful discussions.
PS would like to acknowledge financial support from the
French Agence Nationale de la Recherche through the
contract ANR Mistral.

%


\widetext
\newpage
\begin{center}
\huge{\textbf{Supplementary Material}}
\end{center}

\section*{A. Derivation of the effective two-band model}

In this section we show how to derive the effective two-band Hamiltonian given by Eq.~(5) from the main text. We start by writing the Bogoliubov-de Gennes Hamiltonian for a 2D s-wave superconductor in the Nambu basis  $\left\{c_{\bs k \uparrow}, c_{\bs k \downarrow}, c^\dag_{-\bs k \downarrow}, -c^\dag_{-\bs k \uparrow} \right\}^{\mathrm{T}}$
$$
H_0 = \xi_k \tau_z + \Delta_s\tau_x,
$$
with  $\bs \tau = (\tau_x, \tau_y, \tau_z)$ matrices acting in particle-hole subspace.  The superconducting order parameter is denoted by $\Delta_s$, the spectrum of free electrons is defined as $\xi_k \equiv \frac{k^2}{2m}-\varepsilon_F$, where $\varepsilon_F$ is the Fermi energy. A chain of magnetic impurities with precessing spins deposited on top of the superconductor is given by
$$
H_{imp}(t) = \sum\limits_j {\bs J}_j(t) \cdot \bs \sigma\ \; \delta(\bs r - \bs r_j),
$$
where $\bs \sigma = (\sigma_x, \sigma_y, \sigma_z)$ matrices acting in spin subspace. We assume that the impurities are localized at positions $\bs r_j$, and have precessing spins that are defined as ${\bs J}_j(t) \equiv J \left[\sin\theta \cos (\omega_0 t +\phi_j),\, \sin\theta \sin (\omega_0 t +\phi_j),\, \cos\theta \right]$  with precession frequency $\omega_0$, polar angle $\theta$ as shown in Fig.1 in the main text, and equidistant  individual phase shifts $\phi_j \equiv k_h a j,\, j \in \mathbb{Z}$. In the latter $a$ denotes the spacing between impurities, and $k_h$ is the so-called helix step. Thus the full time-dependent Schr\"odinger equation of the problem reads:
\begin{equation}
i \partial_t \Psi\left(\bs r, t \right) = H_{\rm tot}\Psi\left(\bs r, t \right),
\label{SchEqt}
\end{equation}
where $H_{\rm tot}(t) \equiv H_0+H_{\rm imp}(t)$. Since the Hamiltonian is periodic, $H_{\rm tot}(t+T)=H_{\rm tot}(t)$, with $T=2\pi/\omega_0$ we can make use of the Floquet theorem in order to find the resulting (time-dependent) eigenstates and quasi-energy spectrum. However, below we follow a more elegant path employing a rotating wave transformation. 

\subsection*{1. Rotating wave transformation}

The symmetry of the problem allows us to perform a time-dependent unitary transformation that makes the problem fully static. We can write $\Psi\left(\bs r, t \right) = U(t)\Phi\left(\bs r\right)e^{-iEt} $, with $U(t)=e^{-i\omega_0 t \sigma_z/2}$  so that we obtain the stationary Schr\"odinger equation
\begin{equation}
\left[H_{\rm tot}(0) - B\sigma_z\right]\Phi\left(\bs r\right) = E \Phi\left(\bs r\right),
\label{SchrStatSM}
\end{equation}
where the fictitious magnetic field $B \equiv\omega_0/2$  is perpendicular to the plane of the superconductor, which will be referred to as `driving frequency' hereinafter, and  $E$ is the quasi-energy defined modulo $\omega_0$. Below we rewrite Eq.~(\ref{SchrStatSM}) as
\begin{equation}
\Phi (\bs r) = \sum\limits_j G_0^{eff}(E, \bs r -\bs r_j) V_j^{eff} \Phi(\bs r_j),
\label{SchrStatGlaz}
\end{equation}
where
$$
V_j^{eff} \equiv J \begin{pmatrix} \cos\theta & \sin\theta e^{-i\varphi_j}\\  \sin\theta e^{i\varphi_j} & - \cos\theta \end{pmatrix} \otimes \tau_0
$$
and $G_0^{eff} \equiv  \left[E - \mathcal{H}^{eff}_0 \right]^{-1}$ with $H^{eff}_0 \equiv H_0 - B\sigma_z \otimes \tau_0$. We write the coordinate dependence of the Green's in the following form:
\begin{align*}
G_0^{eff}(E, \bs r) =
	\begin{pmatrix}
		(E + B) X_0^+ + X_1^+ & 0 & \Delta_s X_0^+ & 0 \\
		0 & (E - B) X_0^- + X_1^- & 0 & \Delta_s X_0^- \\
		\Delta_s X_0^+ & 0 & (E + B) X_0^+ - X_1^+ & 0 \\
		0 & \Delta_s X_0^- & 0 & (E - B) X_0^- - X_1^-
	\end{pmatrix}.
\end{align*}
Depending on whether $\bs r \neq \bs 0$ or $\bs r = \bs 0$ we give the expressions for $X_0^\pm$ below:
\begin{align*}
X_0^\pm(r) &= -2\nu\cdot \frac{1}{\omega_\pm} \im K_0 \left[ -i \left(1+i\Omega_\pm \right)p_F r \right], \quad X_0^\pm(\bs 0) = -\pi\nu \frac{1}{\omega_\pm}, \\
X_1^\pm(r) &= -2\nu\cdot \re K_0 \left[ -i \left(1+i\Omega_\pm \right)p_F r \right], \quad X_1^\pm(\bs 0) = 0,
\end{align*}
where $ \omega_\pm = \Omega_\pm v_F p_F \equiv \sqrt{\Delta_s^2 - (E \pm B)^2}$. \\

It is worth noting that the system has two different regimes of driving: gapful when $B < \Delta_s$ and gapless when $B > \Delta_s$. Below we consider the case of a gapful system, and moreover we discuss the case of weak driving $B \ll \Delta_s$.

\subsection*{2. Effective Hamiltonian for the chain in momentum space}
We follow the procedure described in the article by Pientka, Glazman and von Oppen \cite{Pientka2013} and we start with rewriting Eq.~(\ref{SchrStatGlaz}) for $\bs r = \bs r_i$, i.e. for the points where the impurities are localised:
\begin{align}
\left[ \mathbb{I} - G_0^{eff}(E,\bs 0) V^{eff}_i \right] \Phi(\bs r_i) = \sum\limits_{j \neq i} G_0^{eff}(E,\bs r_i - \bs r_j) V^{eff}_j \Phi(\bs r_j).
\end{align}
We consider the so-called deep-dilute regime in which the energies of the impurity-induced states are very close to zero. Therefore, we use an approximation for both left and right side of the equation above, leaving in the left side the diagonal term, linear in E. The equation takes form:
\begin{equation}
\left[ \mathbb{I} - \tilde{G}_0^{eff}(E,\bs 0) V^{eff}_i \right] \Phi(\bs r_i) \approx \sum\limits_{j \neq i} G_0^{eff}(0,\bs r_i - \bs r_j) V^{eff}_j \Phi(\bs r_j),
\label{eqforphis}
\end{equation}
where on the left side we make a Taylor expansion up to terms linear in $E$:
\begin{align}
\tilde{G}_0^{eff}(E,\bs 0) = -\frac{\pi \nu}{\omega} \left[
	\begin{pmatrix}
		B & 0 & \Delta_s & 0 \\
		0 & -B & 0 & \Delta_s \\
		\Delta_s & 0 & B & 0 \\
		0 & \Delta_s & 0 & -B
	\end{pmatrix} +
	\frac{\Delta_s^2}{\omega^2}
	\begin{pmatrix}
		E & 0 & \frac{B}{\Delta_s} E & 0 \\
		0 & E & 0 & \frac{B}{\Delta_s} E \\
		\frac{B}{\Delta_s} E & 0 & E & 0 \\
		0 & \frac{B}{\Delta_s} E & 0 & E
	\end{pmatrix}	
	\right]
\end{align}
with $\omega \equiv \sqrt{\Delta_s^2-B^2}$. Below we will keep only the diagonal terms of the second matrix in Eq.~(17), since $B \ll \Delta_s$. On the right side of Eq.~(\ref{eqforphis}) we have
$$
G_0^{eff}(0, r_{ij}) = \frac{\pi \nu}{\omega}
	\begin{pmatrix}
		B \tX_0 + \omega\tX_1 & 0 & \Delta_s \tX_0 & 0 \\
		0 & -B \tX_0 + \omega\tX_1 & 0 & \Delta_s \tX_0 \\
		\Delta_s \tX_0 & 0 & B \tX_0 - \omega\tX_1 & 0 \\
		0 & \Delta_s \tX_0 & 0 & -B \tX_0 - \omega\tX_1
	\end{pmatrix},
$$
where 
\begin{align}
\tX_0(r_{ij}) &= -\frac{2}{\pi}\, \im \mathrm{K}_0 \left[-i\left(1+i\frac{\omega}{v_F p_F} \right) p_F r_{ij}\right], \\
\tX_1(r_{ij}) &= -\frac{2}{\pi} \, \re \mathrm{K}_0 \left[-i\left(1+i\frac{\omega}{v_F p_F} \right) p_F r_{ij}\right],
\end{align}
and $r_{ij} \equiv |\bs r_i - \bs r_j|=|i-j|a$. Our goal is to write Eq.~(\ref{eqforphis}) as a Shr\"odinger equation, to achieve that we use the following unitary transformation: 
$$
U_i = \diag\left\{ e^{i\varphi_i/2}, e^{-i\varphi_i/2}, e^{i\varphi_i/2}, e^{-i\varphi_i/2} \right\},\quad \tilde{\Phi}(\bs r_i) = U_i \Phi(\bs r_i).
$$
Thus we get
\begin{align}
\left[ \mathbb{I} - \tilde{G}_0^{eff}(E,\bs 0) \cdot V \right] \tilde{\Phi}(\bs r_i) = \sum\limits_{j \neq i} G_0^{eff}(0,r_{ij}) \cdot \begin{pmatrix} e^{i (\varphi_i-\varphi_j)/2} & 0 \\ 0 & e^{-i (\varphi_i-\varphi_j)/2}\end{pmatrix} \otimes \tau_0 \cdot V \cdot \tilde{\Phi}(\bs r_j)
\label{Eqalmost}
\end{align}
with
$$
V \equiv U_i V^{eff}_i U_i^\dag = J 
	R(\theta)
	, \quad \text{where} \; R(\theta)= 
	\begin{pmatrix}
		\cos\theta & \sin\theta \\
		\sin\theta & -\cos\theta 
	\end{pmatrix} \otimes \tau_0
$$
Eq.~(\ref{Eqalmost}) then takes form:
\begin{align}
\sum\limits_j H_{ij} \tilde{\Phi}(\bs r_j) = E \tilde{\Phi}(\bs r_i), \quad i,j \in \overline{1,N}, \quad \text{with } H_{ij} = \begin{cases} h_0, & j = i \\ h_{ij}, & j \neq i \end{cases},
\label{Hij}
\end{align}
where
\begin{align*}
h_0 &= -\frac{\omega^2}{\Delta_s^2} \left[\frac{\omega}{\alpha} R(\theta) + B\, R(2\theta)  + \Delta_s \sigma_0\tau_x   \right]\\
h_{ij} &= +\frac{\omega^2}{\Delta_s^2} R(\theta) \cdot \left\{ \left[B \sigma_z \tau_0 + \Delta_s \sigma_0  \tau_x \right] \tX_0(r_{ij}) + \omega \sigma_0 \tau_z \tX_1(r_{ij}) \right\} \cdot \begin{pmatrix} e^{i (\varphi_i-\varphi_j)/2} & 0 \\ 0 & e^{-i (\varphi_i-\varphi_j)/2}\end{pmatrix} \otimes \tau_0 \cdot R(\theta).
\end{align*}
The phases can be expressed in terms of the helix step $k_h$ and spacing $a$,
$
\varphi_j = k_h x_j = k_h a \cdot j.
$
The system described by Eq.~(\ref{Hij}) is translational-invariant and therefore we can perform a FT to obtain the Hamiltonian in the momentum space, namely:
\begin{align}
\mathcal{H}(k) = \sum\limits_{j} H_{ij} e^{i k r_{ij}} = h_0 + \sum\limits_{j \neq i} h_{ij} e^{i k r_{ij}}
\label{FT}
\end{align} 

The $4 \times 4$ form of the Hamiltonian (\ref{FT}) is not convenient for studying the Shiba band since it takes into account the bands that are very close to the edge of the superconducting gap. Therefore, we perform a unitary transformation 
\begin{align}
\mathcal{U} \equiv \exp \left\{ i\, \frac{\theta + \alpha \frac{B}{\Delta_s}\sin\theta}{2} \sigma_y\right\} \otimes \exp \left\{ i \frac{\pi}{4} \tau_y \right\},
\label{UT}
\end{align}
that in the leading order in $B$ diagonalises $h_0$, namely:
\begin{align*}
\left[ \mathcal{U}h_0\mathcal{U}^\dag \right]_{11} = -\left[\left(1 + \frac{1}{\alpha} \right) \Delta_s +B \cos \theta\right] \\
\left[ \mathcal{U}h_0\mathcal{U}^\dag \right]_{22} =  -\left[\left(1 - \frac{1}{\alpha} \right) \Delta_s - B \cos \theta\right]\\
\left[ \mathcal{U}h_0\mathcal{U}^\dag \right]_{33} = +\left[\left(1 - \frac{1}{\alpha} \right) \Delta_s - B \cos \theta\right]\\
\left[ \mathcal{U}h_0\mathcal{U}^\dag \right]_{44} = +\left[\left(1 + \frac{1}{\alpha} \right) \Delta_s +B \cos \theta\right]
\end{align*}
The $22$ and $33$ elements correspond to the sought-for Shiba band, therefore, we can perform the same transformation for the $h_{ij}$ and extract only the terms $22,23,32,33$, in other words, project our Hamiltonian. Thus below we deal with effective $2 \times 2$ Hamiltonians.\\

Two limiting cases are important for understanding: coherence length $\xi \equiv v_F/\omega$ must be compared to the impurity spacing $a$. The case of $\xi \ll a$ takes into account only the nearest neighbour hopping, therefore in Eq.~(\ref{FT}) we consider only $|j-i| \leqslant 1$, whereas in the case of $ \xi \gg a$ we should take into account all the possible hoppings. In both regimes the Hamiltonian is written in the following form
$$
\mathcal{H}(k) = d_0(k) + \bs{d}(k)\cdot \bs{\Sigma},
$$
where $\bs{\Sigma} = \left(\Sigma_x, \Sigma_y, \Sigma_z \right)$ are Pauli matrices acting in a mixed space defined by the unitary transformation of the initial Nambu basis (see Eq.~(\ref{UT})). The components of the $\bs d$-vector are defined below.
 
\subsubsection*{Short coherence length, $\boldsymbol{\xi \ll a}$}

In this regime we keep only the terms responsible for the nearest neighbour hopping, and we get:
\begin{align*}
& d_0(k) \equiv \tX_0(a) \left[\Delta_s \cos\theta - B(1-\alpha \sin^2\theta) \right] \sin \frac{k_h a}{2} \sin ka \\
& d_x(k) \equiv \tX_1(a)\left(\Delta_s - \alpha B \cos\theta\right) \sin\theta \sin \frac{k_h a}{2} \sin ka \\
& d_y(k) \equiv 0 \\
& d_z(k) \equiv -\left[\left(1 - \frac{1}{\alpha} \right) \Delta_s - B \cos \theta\right] + \tX_0(a) \left(\Delta_s -  B \cos\theta\right) \cos \frac{k_h a}{2} \cos ka
\end{align*}

\subsubsection*{Long coherence length, $\boldsymbol{\xi \gg a}$}
In this case one needs to take into account all the possible hopping terms, therefore we perform the summation in Eq.~(\ref{FT}) up to infinity:
\begin{align}
F_{0,1}(k,s) \equiv 2\sum\limits_{m=1}^\infty \cos\left[ (k + s\frac{k_h}{2})a \cdot m\right] \tilde{X}_{0,1}(a\cdot m)
\label{EQF}
\end{align}
Since it is known that $k_F a \gg 1$, we can use the asymptotic form of the modified Bessel function of the second kind and perform the summations to obtain a closed analytical form for the momentum-space Hamiltonian. It is known that
\begin{align*}
\tilde{X}_{0}(a\cdot m) & \sim -\sqrt{\frac{2}{\pi}} \cdot \frac{\sin(k_F a \cdot m +\pi/4)}{\sqrt{k_F a \cdot m}} e^{-k_S a \cdot m} \\
\tilde{X}_{1}(a\cdot m) & \sim -\sqrt{\frac{2}{\pi}} \cdot \frac{\cos(k_F a \cdot m +\pi/4)}{\sqrt{k_F a \cdot m}} e^{-k_S a \cdot m},
\end{align*}
where $k_S \equiv \omega/v_F$. We perform the summations in Eq.~(\ref{EQF}) using the asymptotic forms above and we get:
\begin{align*}
F_{0}(k,s) = +\sqrt{\frac{2}{\pi k_F a}} \im f(k,s), \quad
F_{1}(k,s) = -\sqrt{\frac{2}{\pi k_F a}} \re f(k,s), 
\end{align*}
with
\begin{align*}
f(k,s) = e^{-i\frac{\pi}{4}} \left[\Li_{\frac{1}{2}}\left( e^{-k_S a + i(k+s k_h/2-k_F)a} \right) + \Li_{\frac{1}{2}}\left( e^{-k_S a - i(k+ s k_h/2+k_F)a} \right) \right],
\end{align*}
where we define the polylogarithm function in a standard way:
$
\Li_n(z) = \sum\limits_{m=1}^\infty z^m/m^n.
$
Finally, the Hamiltonian in the long coherence length regime is defined by:
\begin{align*}
& d_0(k) \equiv \left[\Delta_s \cos\theta - B(1-\alpha \sin^2\theta) \right] \frac{F_0(k,-) - F_0(k,+)}{2}\\
& d_x(k) \equiv \left(\Delta_s - \alpha B \cos\theta\right) \sin\theta \,\frac{F_1(k,-) - F_1(k,+)}{2} \\
& d_y(k) \equiv 0 \\
& d_z(k) \equiv -\left[\left(1 - \frac{1}{\alpha} \right) \Delta_s - B \cos \theta\right] + \left(\Delta_s -  B \cos\theta\right) \frac{F_0(k,-) + F_0(k,+)}{2}  
\end{align*}
Note that the coefficients before the combinations of polylogarithm functions are the same as for the short coherence length regime. It is also worth mentioning that in the functions $F_{0,1}(k,s)$ we can set the factor $e^{-k_S a}$ to unity since $\xi \gg a$.

\section*{B. Winding number calculation}

\begin{figure}[h!]
	\includegraphics*[width = 0.7\columnwidth]{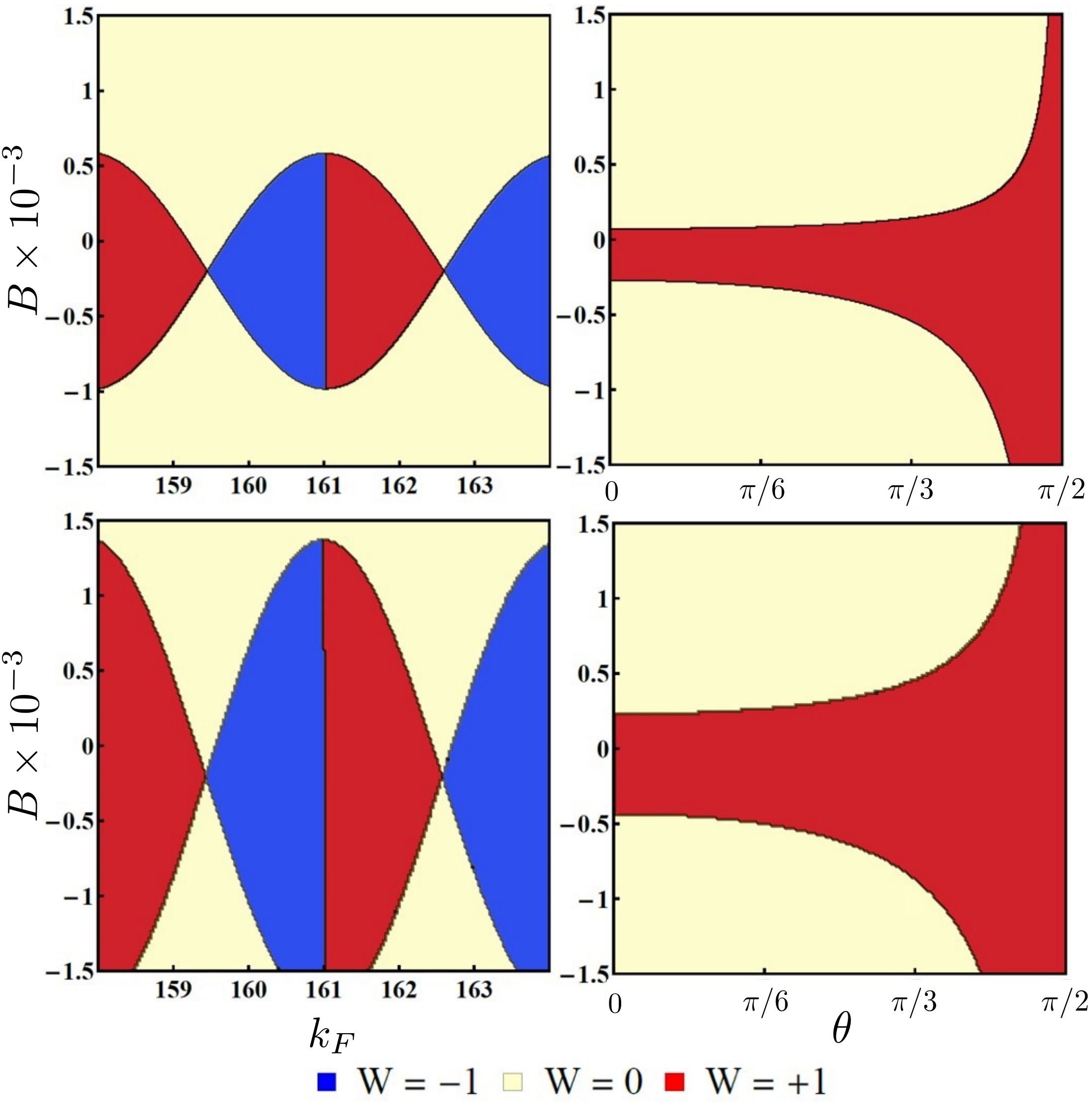}
	\caption{The winding number of the Shiba band for the small and long coherence length regimes (upper and lower rows correspondingly), plotted as functions of the driving frequency $B$ (vertical axis) versus the Fermi momentum $k_F$ (left column, $\theta = \pi/3 $) and polar angle $\theta$ (right column, $k_F = 159$). We set $k_h=\pi/4, v_F = 0.2, \Delta_s = 1, a=1, \alpha = 0.9999$.}
	\label{windingfigure}
\end{figure}

The $d_0$ component of the Hamiltonian does not affect its topological properties. The winding number can be computed employing a standard formula
$$
W = \frac{1}{2\pi} \int\limits_{-\pi}^{\pi} \negthickspace dk\, \frac{d_x(k)d'_z(k)-d'_x(k)d_z(k)}{d^2_x(k) +  d^2_z(k)}.
$$
The results are plotted in Figure \ref{windingfigure} for short and long coherence length regimes as functions of driving $B$ versus either Fermi momentum $k_F$ and polar angle $\theta$. The black lines defining where the winding number changes were utilized in Fig.2 in the main text to facilitate distinguishing between different phases.

\begin{figure}[t]
	\includegraphics*[width = 0.8\columnwidth]{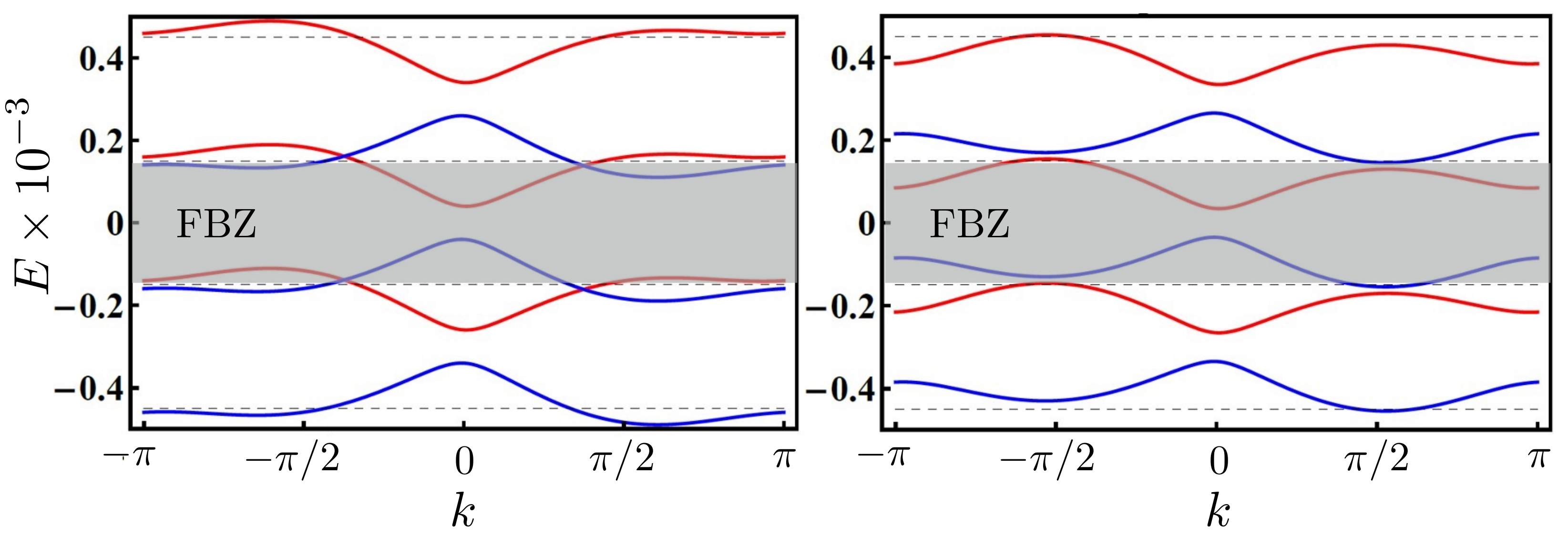}
	\caption{We plot Shiba bands as functions of energy on the left panel and quasi-energy on the right, versus quasi-momentum in the first Brillouin zone. We set $\theta = \pi/3, k_h=\pi/4, k_F=159.3, v_F = 0.2, \Delta_s = 1, a=1, \alpha = 0.9999$ for both panels. On the left $B=0$, whereas on the right $B = -0.15 \cdot 10^{-3}$.}
	\label{anticrossings}
\end{figure}

\section*{C. Frequency domain description of the Shiba bands}

In this section we describe the dynamical situation in the frequency domain, and the resulting Floquet band structure. Given a periodic Hamilton $H(t+T)=H(t)$, with some period $T=2\pi/\omega_0$, one can write the time dependent Schroedinger equation:
\begin{equation}
i\frac{\partial\psi(t)}{\partial t}=H(t)\psi(t)\,,
\end{equation}   
and one can utilize the Floquet theorem to subsequently write:
\begin{equation}
\psi(t)=e^{-iEt}\sum_{m=-\infty}^\infty\phi_me^{im\omega_0t}\,,
\end{equation}
with $\phi_m$ depending on various system parameters, but not $t$. These coefficients (or wave functions) satisfy the following time-independent eigenvalue equation:
\begin{equation}
\sum_{m'}H_{mm'}\phi_{m'}=E\phi_m\,,
\end{equation} 
where
\begin{equation}
H_{mm'}=\omega_0\delta_{mm'}+\frac{1}{T}\int_0^Tdte^{i(m-m')\omega_0t}H(t)\,.
\label{spectrum}
\end{equation}
We see that the above equation has solutions has solutions for $-\infty<E<\infty$. However,  if $E$ is an
eigenvalue of $\eqref{spectrum}$  corresponding to the eigenstate $m$
with amplitudes $\phi_{m}$, then $\tilde{E}=E+s\omega_0$, where $s$
is any integer, is also an eigenvalue corresponding to an
$\tilde{\phi}_{m+s}$  eigenstate with amplitudes given by $\tilde{\phi}_{m+s}=\phi_m$. This means that all of these solutions correspond to the same time-dependent solution of the Schrodinger equation. Therefore, the Floquet states are uniquely described by quasi-energies in the ``first quasi-energy Brillouin zone" (FBZ), $-\pi/T<E<\pi/T$. Equation \eqref{spectrum} is the temporal analogue of a repeated zone scheme for usual band structure calculations. Let us consider our model of a driven spin helix coupled to a superconductor. To ease the discussion, it is instructive to write this Hamiltonian as follows:
\begin{align}
H_{\rm tot}(t)&=H_{0}+\sum_{j=1}^N\left[J_j^z\sigma_z+J_j^+\sigma_-e^{-i\omega_0t}+J_j^-\sigma_+e^{i\omega_0t}\right]\delta(\bm{r}-\bm{r}_j)\,,
\end{align}
where $J_j^{z}=J_0\cos{\theta}$, and $J_{j}^{\pm}=J_0\sin{\theta}e^{\pm i\phi_j}$, and $H_0$ is the superconducting Hamiltonian that was defined before.  Note that $[H_0,\sigma_z]=0$, and that we can identify the $J_{j}^{+(-)}$ processes with emission (absorption of a  photon, or any bosonic quanta). Note also that the functions $\phi_m$ depend, among other things, on the spin degree of freedom.  With that identification, we immediately can establish that the resulting Hamiltonian in Eq.~\eqref{spectrum} can be decomposed in blocks of $2\times2$ in the spin$\otimes$photon space, or $\{\phi_{m\downarrow},\phi_{m+1\uparrow}\}$ constitute a closed basis (does not couple to other bands). More specifically, we obtain the following matrix:
\begin{align}
H_{\rm mm-1}=\left(
\begin{array}{cc}
H_0+m\omega_0-\sum_j J_j^z\delta(\bm{r}-\bm{r}_j) & \sum_{j}J_{j}^+\delta(\bm{r}-\bm{r}_j)\\
\sum_{j}J_{j}^-\delta(\bm{r}-\bm{r}_j) & H_0+(m-1)\omega_0+\sum_j J_j^z\delta(\bm{r}-\bm{r}_j)
\end{array}
\right)\,,
\label{effective}
\end{align} 
which can be casted in terms of a ``pseudospin" $\tilde{\bm{\sigma}}$ as follows:
\begin{equation}
H_{m,\downarrow}^{m-1,\uparrow}=(m+1/2)\omega_0+H_0-(\omega_0/2)\tilde{\sigma}_z+\sum_j \bm{J}_j\cdot\tilde{\bm{\sigma}}\,\delta(\bm{r}-\bm{r}_j)\,.
\end{equation} 
Note that from the perspective of the superconductor, $\bm{\sigma}$ and $\tilde{\bm{\sigma}}$ act exactly in the same fashion. We can now easily interpret our results: in the extended zone scheme, the initial bands only couple the same $m's$  in the absence of driving.  However, once the driving is tuned on, the bands with adjacent $m$'s and opposite spins get mixed. Moreover, each pair of such bands are shifted in energy by $\omega_0(m+1/2)$. To highlight this behavior, it worth doing the following comparison: the static spectrum, in the absence of driving, but presented in the extended zone scheme for a given frequency $\omega_0$ (copies of the spectrum shifted by $\omega_0(m+1/2)$, but not interacting with each other for different $m$'s), and the Floquet spectrum for the driven system, presented again in the extended zone scheme. We see that indeed, in the driven case, there is level crossing between adjacent $m$'s, with a splitting that can be evaluated from solving Eq~\eqref{effective}. Since the spectrum is composed of separated blocks that emulate the static Hamiltonian, indifferent of the periodic or open boundary conditions, the entire edge structure and topology is given by these (shifted, but equivalent) blocks.   

On the left panel in Figure \ref{anticrossings} we plot the Shiba band energies versus quasi-momentum in the absence of driving, showing also the Floquet-like band structure (bands are replicated artificially, using the value of driving for the right panel, since there is no driving on the left one). When we turn on the driving we see that the crossings of the left panel become the anticrossings on the right panel, where we plot quasi-energies (Floquet bands) of Shiba band versus quasi-momentum. This qualitative analysis is analogous to that carried out by M. Rudner et al. (see Fig. 5 of PRX \textbf{3}, 031005), and allows to see how the topological phase transition occurs in this system subject to periodic driving.

\section*{D. Shiba band gap at $\bs{E=B}$}

\begin{figure}[h!]
	\includegraphics*[width = 0.7\columnwidth]{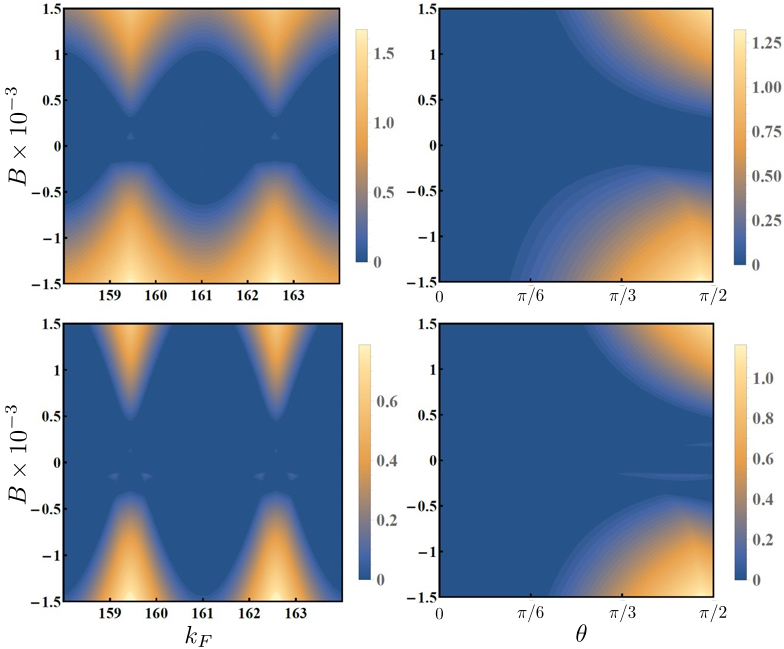}
	\caption{The gap around quasienergy $E=B$ of the Shiba band for the small and large coherence length regimes (first and second rows respectively), plotted as functions of the driving frequency $B$ and the Fermi momentum $k_F$ (precession angle $\theta$) in the left (right) column. We set $k_h=\pi/4, v_F = 0.2, \Delta_s = 1, a = 1, \alpha = 0.9999$. The Fermi momentum $k_F=159$ in the left column and the polar angle $\theta=\pi/3$ in the right column.}
	\label{gapatB}
\end{figure}

In Fig.2 in the main text we presented only the gap value of the Shiba band taken at quasi-energy $E=0$. For completeness we present also in Figure \ref{gapatB} the gap at quasi-energy $E=B$. As discussed in the main text, no Majorana fermions emerge at the latter quasi-energy value.

\section*{E. Circuit implementation of a precessing helical texture}

\begin{figure}[t]
	\includegraphics*[width = 0.7\columnwidth]{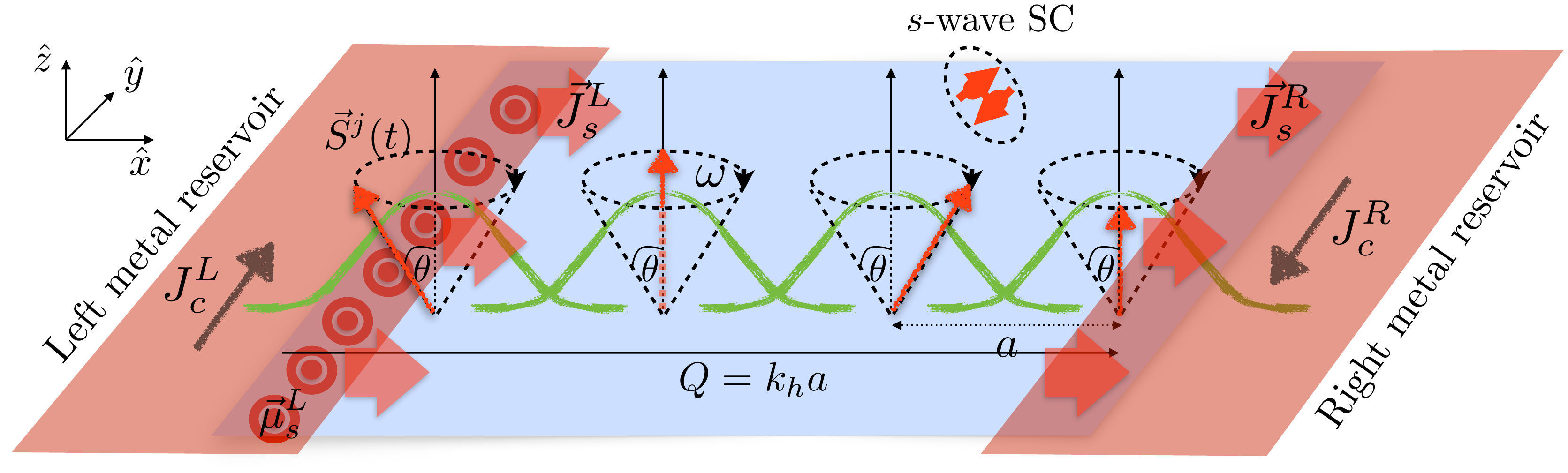}
	\caption{Normal-metal|Magnetic Texture|normal-metal heterostructure that can be
used to excite in a controlled fashion the precession of the chain of magnetic impurities inserted in an $s$-wave superconductor.  A charge current
$J_c^L$ in the left lead (red) gives rise to a spin accumulation $\vec{\mu}_s^L\equiv \vec{e}_z\mu_s^L$ at the
interface via the spin Hall effect, and the spin current pumped into
the right reservoir, $\vec{J}_S^R$, generates a transverse charge current $J_{c}^R$ through
the inverse spin Hall effect. The spin current from the left ($\vec{J}_s^L$) to the right lead ($\vec{J}_s^R$)
 is established dynamically via the precession of the magnetic (helical) texture. }
	\label{Transport}
\end{figure}

In this section we present a simplified version of the implementation of a controlled dynamical helical texture in a circuit model. In Fig.~\ref{Transport} we present a sketch of our proposal: a chain of magnetic impurities on top of an $s$-wave superconductor forming spin helix (texture) of pitch $k_h$ and an angle $\theta$ is being sandwiched between to metallic leads with strong spin-orbit interaction. Passing a charge current $J_c^L$ through the left metal gives rise to a spin accumulation $\bm{\mu}_s^L=\bm{e}_z\mu_s^L$ at the edge via the spin Hall effect \cite{SinovaRMP15}. This spin bias, being non-collinear with the local magnetization on the left, gives rise to a local torque that brings the texture into precession. Below we describe a simplified version of the model that describes quantitatively the induced precession of the texture, assuming the texture is rigid and that the entire spin is carried by the texture only. Moreover, for simplicity, we assume no bulk Gilbert damping of the (precessing) magnetic texture, although such a component is easily accommodated in our theory.  Our theoretical modeling follows closely the ideas developed in Ref.~\cite{TakeiPRL14} for describing superfluid spin flow in planar ferromagnets. Following Ref. \cite{TserkovnyakPRL02}, the spin current exchanged with the lead $r=L,R$ by the helix can be expressed as:
\begin{align}
\bm{J}_s^r=\frac{1}{4\pi}\left[\mathcal{R}g_{\uparrow\downarrow}^{r}\bm{m}\times+\mathcal{I}g_{\uparrow\downarrow}^{r}\right](\tilde{\bm{\mu}}_s^r\times\bm{m})\,,
\end{align}      
where $g_{\uparrow\downarrow}^{r}\equiv\mathcal{R}g_{\uparrow\downarrow}^r+i \mathcal{I}g_{\uparrow\downarrow}^r$ is the so called spin mixing conductance, which quantifies the interface $r=L,R$ and which can be in general complex \cite{TserkovnyakRMP05}, and $\bm{m}\equiv \bm{S}/S_0$. Also, $\tilde{\bm{\mu}}_s^r\equiv\bm{\mu}_s^r-\hbar \bm{m}\times\dot{\bm{m}}$, namely in the dynamical case the applied spin bias (or spin torque) is reduced effectively by the spin pumping contribution \cite{TserkovnyakPRL02}. Next we note that the spin texture posses  $U(1)$ symmetry, namely it is invariant with respect to rotations around the $z$ axis. That implies that the spin current along this direction is conserved, and thus we focus on the $z$ component only. Also, we assume that for a rigid texture (helix)  the spin precession satisfies:
\begin{equation}
\bm{m}\equiv\bm{m}(t)=\left[\cos{(\omega_0t+\phi)}\sin{\theta},\sin{(\omega_0t+\phi)}\sin{\theta}, \cos{\theta}\right]\,,
\end{equation}
with $\phi$ arbitrary. With this, the spin currents along $z$ at each end (lead)  is found as follows:
\begin{equation}
J_{s,z}^{L,R}=\pm\frac{\mathcal{R}g_{\uparrow\downarrow}^{L,R}}{4\pi}(\mu_s^{L,R}-\omega_0)\sin^2{\theta}\,,
\end{equation}  
where we used the $-$ sign for the right spin current to emphasize that this is the current injected into the lead and not the flow into the texture (which is just the opposite). Since we assume no dissipation mechanisms in the bulk we have that $J_{s,z}^L=J_{s,z}^R$, which gives for the precession frequency the following expression:
\begin{equation}
\omega_0=\frac{\mathcal{R}g_{\uparrow\downarrow}^{L}\mu_s^L+\mathcal{R}g_{\uparrow\downarrow}^{R}\mu_s^R}{\mathcal{R}g_{\uparrow\downarrow}^{L}+\mathcal{R}g_{\uparrow\downarrow}^{R}}\,.
\end{equation}  
We thus see that by controlling the spin biases on the two ends, via the spin Hall,  effect the precession frequency can be tuned at will. Moreover, by tuning the biases such that $\mathcal{R}g_{\uparrow\downarrow}^{L}\mu_s^L=-\mathcal{R}g_{\uparrow\downarrow}^{R}\mu_s^R$, the precession frequency vanishes.  More complicated models (including, for example, dissipation, change in the cone angle, electrons, magnons, etc) will alter this conclusion, but the general picture should still hold true. For completeness, we also provide with the resulting expression for the spin current at either of the ends:
\begin{equation}
J_{s,z}^L=\frac{1}{4\pi}\frac{\mathcal{R}g_{\uparrow\downarrow}^{L}\mathcal{R}g_{\uparrow\downarrow}^{R}(\mu_s^L-\mu_s^R)}{\mathcal{R}g_{\uparrow\downarrow}^{L}+\mathcal{R}g_{\uparrow\downarrow}^{R}}\sin^2{\theta}\,,
\end{equation}
which vanishes any time one of the leads is disconnected. However, it does not necessary vanish in the case when $\omega_0$ vanishes (see condition above).  We note in passing that relaxing the condition of a rigid texture should allow, as stressed in the main text, to control non only the frequency, but also the pitch $k_h$. For example, if the initial impurity spins are lying in-plane, but forming no helix in the absence of the leads, once subjected to spin biases will give rise to a dynamical texture such that the spin current is transported from the left to right lead. In a long wavelength description, the Hamiltonian describing such a setup reads \cite{SoninAdvMat10,TakeiPRL14}:
\begin{equation}
H_F=\int dx[A(\nabla\bm{m}(x))^2+Km_z^2]/2\,,
\end{equation}
where $A$ and $K$ stand for the exchange stiffness and the anisotropy, respectively, where $\bm{m}=(\sqrt{1-m_z^2}\cos{\phi},\sqrt{1-m_z^2}\sin{\phi}, m_z)$. For $m_z\ll1$, it was shown that the spin current (along $z$) at some position $x$ in the chain between the two leads reads:
\begin{equation}
J_{s,z}(x)=-A\nabla\phi(x)\,,
\end{equation} 
which, in the absence of dissipation, should coincide with the spin currents at the left lead  $J_{s,z}^L$. In such a case, we can readily see that $\phi(x,t)=\phi(0,t)-(J_{s,z}^L/A)x$, with $\phi(0,t)=\omega_0t$. The resulting pitch of the texture (assuming, in the lattice model  $a=1$) can be read as
\begin{equation}
k_h=2\pi A/J_{s,z}^L\,.
\end{equation}
By inspecting the above conditions for the precession frequency and for the spin current, we see that by tuning the spin biases we can tune independently the frequency $\omega_0$ and the pitch $k_h$ (we can even reduce the chain to the static case). Thus, such an implementation is extremely versatile for implementing our proposed dynamical model.

\end{document}